\documentclass[twocolumn, superscriptaddress]{revtex4-2}
\usepackage[english]{babel}
\usepackage[T1]{fontenc}
\usepackage{amsmath,amssymb}
\usepackage{graphicx}

\usepackage{xcolor}
\usepackage{bm}      
\usepackage{physics}  
\usepackage[normalem]{ulem}
\usepackage{xurl}
\usepackage{hyperref}
\hypersetup{
    colorlinks=true,
    linkcolor=blue,
    citecolor=blue,
    urlcolor=blue
}
\begin{document}
\title{Resonant Amplification of Demographic Noise in Stochastic Common-Pool Resource Dynamics}

\author{Susanna C. Caruso}
\affiliation{Dipartimento di Fisica ed Astronomia, Università di Catania and INFN, Catania I-95123, Italy}
\author{Fabio Menegazzo}
\affiliation{Laboratory of Interdisciplinary Physics, Department of Physics and Astronomy, University of Padova, Padova, Italy}
\author{Vito Latora}
\affiliation{Dipartimento di Fisica ed Astronomia, Università di Catania and INFN, Catania I-95123, Italy}
\affiliation{School of Mathematical Sciences, Queen Mary University of London, London E1 4NS, United Kingdom}
\affiliation{Complexity Science Hub Vienna, A-1080 Vienna, Austria}
\author {Samir Suweis}
\affiliation{Laboratory of Interdisciplinary Physics, Department of Physics and Astronomy, University of Padova, Padova, Italy}
\affiliation{INFN, Sezione di Padova,via Marzolo 8, 35131 Padova, Italy}

\begin{abstract}

Common-pool resources are vulnerable to overexploitation, yet many social–ecological systems exhibit persistent oscillations.  We introduce a stochastic eco-evolutionary model in which harvesting strategies and resource availability shape one another. In the infinite population limit the system approaches a stable state through damped oscillations. Demographic noise continually excites these modes, generating instead robust macroscopic quasi-cycles, whose frequencies depend on the noise ratio, the relative timescale between strategy and resource, and the harvesting pressure. The model reproduces the distinct frequency peaks, observed in empirical global fish-landings data for high- and low-income countries, showing that persistent environmental fluctuations can emerge naturally from demographic noise.
\end{abstract}
\maketitle

Common-pool resources (CPRs) are finite and  non-excludable shared resources. 
These characteristics make them vulnerable to overexploitation, potentially leading to the so-called ``tragedy of the commons''   \cite{hardin_tragedy_1968}. 
As highlighted by Kobeissi et al. \cite{kobeissi_tragedy_2024} and others \cite{feeny_tragedy_1990, ostrom_tragedy_2008}, in CPRs  individual incentives promote overexploitation over coordinated collective benefit, ultimately driving resource depletion. 
However, this tragedy is not inevitable. The seminal works by Nobel Prize laureate Elinor Ostrom \cite{ostrom_challenge_2008} alongside recent theoretical and behavioral experiments \cite{tu_emergence_2023}, demonstrates that introducing shared objectives can override short-term self-interest and sustain cooperation.
Real-world CPR dynamics are governed by complex human-ecological feedbacks. Classical examples includes the alternating dominance of sardine and anchovy stocks observed in the Pacific \citep{chavez_anchovies_2003,sugihara_detecting_2012}, as well as historical fishery collapses, where intensive harvesting interacts with ecological cycles \citep{hutchings_collapse_2000,myers_population_1995}.
When resource-use strategies depend on both social interactions and environmental states, their coevolution is naturally captured by coupling evolutionary game theory (EGT) with resource dynamics \cite{weitz_oscillating_2016,tilman_evolutionary_2020,tu_emergence_2023,ito_complete_2024,hauert_asymmetric_2019}. The resulting, so-called eco-evolutionary (ECO-EVO) models comprise three key components: game structure, resource dynamics, and environment-dependent strategy adoption. Crucially, in ECO-EVO models the payoff matrix is no longer static; incentives to cooperate (preserve) or defect (overexploit) shift dynamically with resource availability. A standard formulation represents these state-dependent payoffs as a linear combination of games corresponding to depleted and abundant environments \cite{weitz_oscillating_2016, tilman_evolutionary_2020,ito_complete_2024}. Despite their simplicity, ECO-EVO 
models exhibit a rich dynamical repertoire, including bistability, strategy coexistence, and sustained oscillations characterizing recurrent tragedy-of-the-commons cycles.   
Most ECO-EVO models couple deterministic replicator equations with resource dynamics. However, empirical observables like fish productivity \cite{jonzen_theory_2002,anderson_why_2008}, forests growth and supply \cite{yoshimoto_new_1996} and freshwater systems \cite{tabi_species_2024} indicates that deterministic dynamics alone cannot account for observed resource dynamics. 
While previous works have highlighted resonantly amplified quasi-cycles driven by demographic noise \cite{bladon_evolutionary_2010} or explored extrinsic environmental fluctuations \cite{bodin_eco-evolutionary_2026}, they decoupled stochasticity from the bidirectional feedback between strategy evolution and resource dynamics.
Recently, this feedback has been integrated into numerical models exploring knowledge feedback and network topology \cite{saeedian_modelling_2024}, and into system-level dynamical frameworks to derive analytical conditions for stochastic asymptotic stability \cite{zhao_stochastic_2026}. However, these approaches formulate stochasticity directly at the macroscopic scale by introducing independent noise processes. Consequently, the role of intrinsic, finite-size stochasticity within this bidirectional feedback system remains largely unexplored.
Here, we bridge this gap, by introducing a stochastic ECO-EVO model of common-pool resource dynamics in which, rather than postulating macroscopic noise, strategy updates, resource harvesting, and resource regeneration are all treated as coupled stochastic processes at the individual level. 
Starting from a microscopic description, we systematically derive the macroscopic (Langevin) dynamics of the coupled agent–resource system, allowing us to move beyond purely numerical treatments and quantify  analytically how demographic noise reshapes the stability and sustainability of the coupled social-ecological system. 

We also show that our ECO-EVO model reproduces the variation in oscillation frequencies observed in fish-landings data across countries, linking this variation to differences in harvesting pressure and adaptation timescales between high- and low-income countries.

\medskip 
\subsection{Stochastic eco-evolutionary model.}~
Consider a population of $N$ agents, where each individual adopts one of two strategies: cooperation, $c$, or defection $d$. These behaviors determine how individuals interact with one another and how they exploit a shared, renewable resource, which coevolves with the population of strategies. In line with standard treatments in chemical kinetics \cite{kampen_stochastic_2011}, we model the resource as a discrete quantity to facilitate the stochastic description that follows. Hence, at any given time, the system is defined by a population of $C$ cooperators and $D$ defectors, where $C+D=N$, together with a resource level $R \in [0,1,2, \dots K]$, where $K$ denotes the carrying capacity. To quantify how, in our individual-based model, different interactions impact the resource, we introduce the harvest-interaction matrix $E =\{(e_{cc},\,e_{cd}),(e_{dc}, \, e_{dd})$, whose entry $e_{ij}$ represents the per-capita extraction rate exerted by an individual of strategy $i\in\{c,d\}$ when interacting with one of strategy $j\in\{c,d\}$. 
Following \citet{tu_emergence_2023}, cooperators are defined by extraction rates $0 < e_{cc} < b$, with $b$ being the intrinsic resource regeneration rate, 
which allow the resource to renew faster than it is consumed. 
Defectors, by contrast, satisfy $e_{dd} > b$, so the resource becomes extinct at stationarity if all agents choose to defect.
The total extraction rates contributed by a cooperator and a defector in a well-mixed population of size $N$ are, respectively:  
\begin{equation}
\begin{aligned}
h_c(R,C,D) &= \frac{R}{N(N-1)}\Bigl[e_{cc}(C-1)+ e_{cd}D\Bigr], \\
h_d(R,C,D) &= \frac{R}{N(N-1)}\Bigl[e_{dc}C + e_{dd}(D-1)\Bigr],
\end{aligned}
\end{equation}
where we have assumed that the harvested units scale with the available resource $R$, and the normalization $1/[N(N-1)]$ reflects the even distribution of $R$ across all pairwise interactions, weighted by the extraction coefficients $e_{ij}$. The resource depletion is then modeled as a stochastic process occurring at a total rate $h_{c}C+h_{d}D,$ $
\quad r \xrightarrow[]{\,h_{c}C+h_{d}D\,} v$ where the lower-case letters denote the type of species occupying a given site: $r$ represents a unit of  resource, and $v$ represents an empty (available) site, as would naturally arise in a spatial formulation.
Resource regeneration is described as a microscopic birth process occurring at rate $b $, $\quad r+v \xrightarrow[]{\,b\,} r+r$. Formally, this leads naturally to logistic growth. The finite carrying capacity $K$ prevents unbounded growth, as the resource can only occupy the $K-r$ currently empty sites.

\begin{figure}    \centering\includegraphics[width=1\linewidth]{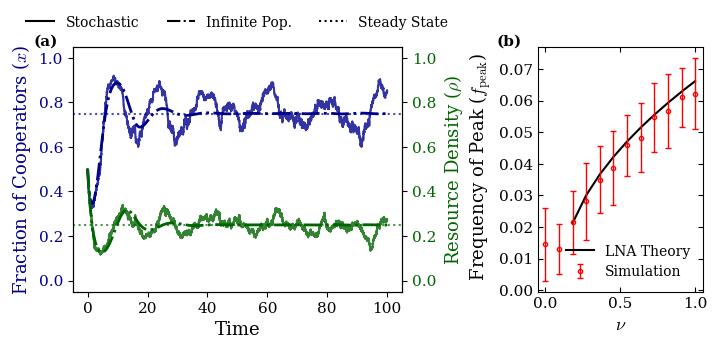}
    \caption{ (a) A single trajectory  of the stochastic ECO-EVO model with $K = 1000$, $N=200$, and  
    $e_{cc} = e_{cd} = 0.5$, $e_{dd} = e_{dc} = 1.5$, $b = \nu = 1$, numerically simulated (solid lines), reveals sustained quasi-cycles in 
    the cooperator fraction $x$ and the resource level $\rho$.
In the infinite-population limit of Eq.~(\ref{eq:langevin}) the system undergoes damped oscillations (dash-dotted lines) around the interior fixed point given by \eqref{eq:interior_fixedpoint} (horizontal dotted line).
(b) Peak frequency $f_{\rm peak}$ of the power spectrum of $\rho$ as a function of the timescale $\nu$. The solid line is the LNA prediction; red symbols show the mean over $n=100$ Gillespie runs of length $t_{\max}=500$ (after discarding an initial transient), with error bars indicating one standard deviation. Other parameters as in (a).} 
    \label{fig:Trajectories}
\end{figure}

Strategy updates occur also stochastically and depend both on agents’ interactions and on the current resource level, thereby incorporating the environmental feedback. Following the knowledge-feedback mechanism introduced in \cite{saeedian_modelling_2024}, combined with imitation dynamics, agents are more likely to switch to defection when others are overexploiting the resource and the resource stock is abundant. Conversely, resource scarcity together with widespread cooperation increases the propensity to adopt a cooperative strategy. 
Specifically, we assume that the rate of switching to defection is proportional to the resource stock $R$, whereas the rate of switching to cooperation increases when resources are scarce and most agents contribute to their preservation.
The two strategy-update processes are therefore $c\xrightarrow[]{\,\nu R' \frac{D}{N}\,} d$ and $d \xrightarrow[]{\,\nu (1-R')^{\tiny{+}}
\frac{C}{N}\,} c$ where $\nu$ sets the strategy-update timescale, $R' = R/[K(1-e_{cc}/b)]$ is the level of resource $R$ rescaled by the effective carrying capacity \cite{saeedian_modelling_2024}, and $(u)^{\tiny{+}} \equiv \max(0,u)$ prevents negative rates for $R'>1$ (see SM). 
The dynamics of the 
stochastic ECO-EVO model can be investigated numerically using the Gillespie algorithm \cite{gillespie_general_1976}. 
Fig.~\ref{fig:Trajectories} (a) reports the fraction of cooperators $x$ and the resource level $\rho$ (solid line) as functions of time for a single realization of the process. Results clearly reveal sustained oscillations with a frequency whose $\nu$ dependence is shown Fig. \ref{fig:Trajectories}(b)(red symbols).  

\medskip
\subsection{Analytical description.}~
To uncover the origin of these oscillations, we derive an analytical description of our model within the master-equation framework \cite{kampen_stochastic_2011}.
From the microscopic processes of the model we first construct nonnegative transition rates between states $(C,D,R)$, with transitions outside the physical domain $0\leq C,D\leq N$ and $0\leq R\leq K$ forbidden. These rates determine the master equation for the state probability $P(C,D,R,t)$ of the system to have $C$ cooperators, $D$ defectors and $R$ resource level at time $t$.
Introducing the cooperator fraction $x=C/N$ and the scaled resource level $\rho=R/K$, the diffusion approximation yields the 
Itô-Langevin equations
\begin{equation}
\begin{aligned}
\frac{d\rho}{dt}
&= \rho\bigl[b(1-\rho)-g(x)\bigr]
 + \sqrt{\frac{B_{\rho}(\rho,x)}{K}}\,\eta_{\rho}(t),\\
\frac{dx}{dt}
&= \nu x(1-x)(1-2\rho')
 + \sqrt{\frac{B_x(\rho,x)}{N}}\,\eta_x(t),
\end{aligned}
\label{eq:langevin}
\end{equation}
where $\nu$ control the relative timescale of strategy updates and resource dynamics, $\rho'=\rho/(1-e_{cc}/b)$, and
\[
g(x)=e_{cc}x^2+(e_{cd}+e_{dc})x(1-x)+e_{dd}(1-x)^2
\]
is the harvest pressure. In the diffusion terms, the coefficients obtained from the second jump moments of the microscopic dynamics, are
\begin{equation*}
\begin{aligned}
B_{\rho}(\rho,x)&=\rho\bigl[b(1-\rho)+g(x)\bigr],\\
B_x(\rho,x)&=\nu x(1-x).
\end{aligned}
\end{equation*}
while $\eta_{\rho}$ and $\eta_x$ are independent Gaussian white noises with zero mean and correlations
$\langle\eta_i(t)\eta_j(t')\rangle=\delta_{ij}\delta(t-t')$.
Resource and population fluctuations therefore scale as $K^{-1/2}$ and $N^{-1/2}$, respectively.
Eq. (\ref{eq:langevin}) holds for $\rho' \le 1$; the full expressions for $\rho'>1$ and derivation details are given in the SM

\medskip 
\subsection{Infinite-population limit.}
\begin{figure}
    \centering\includegraphics[width=1\linewidth]{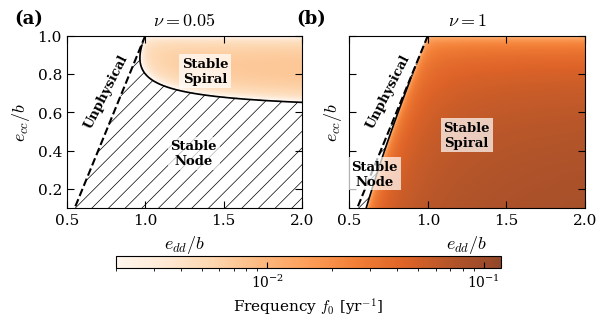}
    \caption{Characterization of the interior fixed point in the infinite-population limit of Eq.~(\ref{eq:langevin}). Shown are the cases of two relative timescales: (a) $\nu = 0.05$ and (b) $\nu = 1$. Axes denote the rescaled extraction rates $e_{dd}/b$ and $e_{cc}/b$, with cross-interactions fixed to $e_{cc} = e_{cd}$ and $e_{dd} = e_{dc}$. The dashed line $e_{dd}=(b + e_{cc})/2$ separates the physically admissible region ($x^* \ge 0$) from the nonphysical regime. The solid line marks the transition to complex eigenvalues. Within the stable spiral regime, the heatmap encodes $f_0$ the frequency of damped oscillations.}
    \label{fig:Fig2}
\end{figure}

~
In the large-system limit ($K,N\to\infty$ at fixed $K/N$), fluctuations vanish, and Eq.~(\ref{eq:langevin}) reduces to two coupled deterministic equations. These admit three boundary fixed points and one interior equilibrium in the admissible state space $\Omega=[0,1]\times[0,1]$. The boundary fixed points are:
\begin{equation}
(\rho^*,x^*)\in
\left\{(0,0),\;(0,1),\;\left(1-\frac{e_{cc}}{b},1\right)\right\}.
\end{equation}
The first two correspond to resource extinction under pure defection and pure cooperation, whereas the third describes a fully cooperative population sustaining the resource ($e_{cc} < b$). Because $e_{dd} > b$, the corresponding pure-defection state $\left(1-e_{dd}/b,\, 0\right)$ is unphysical ($\rho^* < 0$), leaving $(\rho^*=0, x^*=0)$ as the sole equilibrium on the $x=0$ boundary (the ``tragedy of the commons'').
The interior equilibrium is
\begin{equation}
(\rho^*,x^*)=
\left(\frac{1}{2}\left(1-\frac{e_{cc}}{b}\right),\tilde x\right),
\label{eq:interior_fixedpoint}
\end{equation}
where $\tilde x\in(0,1)$ satisfies
\begin{equation}
\frac{1}{2}\left(1-\frac{e_{cc}}{b}\right)
=
1-\frac{g(\tilde x)}{b}.
\end{equation}
Because defectors increase harvesting ($g(x) \ge e_{cc}$), the resource decreases whenever $\rho' \ge 1$ ($\dot\rho \le 0$). Trajectories are thus confined to $\rho' \le 1$, and since the interior equilibrium sits at $\rho'^* = 1/2$ and excursions into $\rho' > 1$ are exponentially rare (see SM).
At $(\rho^*,x^*)$, the Jacobian has strictly negative trace,
$\tau=-b\rho^*<0$, excluding both a Hopf bifurcation and a center. More generally, the Bendixson--Dulac theorem rules out periodic orbits in the interior of $\Omega$ for the deterministic dynamics \cite{bendixson_sur_1901,gasull_effectiveness_2021} (see SM).
The equilibrium is therefore always stable, and approached either monotonically or through damped oscillations,  depending on whether $\tau^2 > 4\Delta$ (stable node) or $\tau^2 < 4\Delta$ (stable spiral), where $\Delta$ is the determinant of the Jacobian.
This spiral relaxation is illustrated by the solutions of the coupled deterministic equations, reported in Fig.~\ref{fig:Trajectories} (a)(dash-dotted lines). Fig.~\ref{fig:Fig2} further illustrates how the nature of the interior fixed point changes as a function of the extraction rate 
of two defectors, $e_{dd}$, and of that of two cooperators, $e_{cc}$.
Without loss of generality, we assume homogeneous interaction strengths:  
$e_{dc}=e_{dd}$ and $e_{cd}=e_{cc}$ \cite{saeedian_modelling_2024}. Two different values of the relative timescale $\nu$ are considered. 
When strategy update is slow compared to the 
resource dynamics
($\nu \ll 1$, e.g., $\nu=0.05$), the parameter space supporting a stable spiral is exceedingly narrow (Fig.~\ref{fig:Fig2}(a)).
In this regime, the resource rapidly approaches its stationary value, while strategies evolve too slowly to induce any significant feedback. 

In contrast, as  $\nu$
increases (e.g., $\nu=1$), strategies react more rapidly to variations in resource abundance.
This has two effects: it significantly expands the parameter region where the equilibrium is a stable spiral, and also increases the associated oscillation frequency  (Fig.~\ref{fig:Fig2}(b)).
Such frequency, $f_0= \frac{1}{2\pi}\sqrt{\Delta - \tau^2 / 4 }$,  
is determined by the imaginary part of the Jacobian eigenvalues, and it is reported by the color gradient in Fig \ref{fig:Fig2}.

\medskip
\subsection{Finite population.}~
Having ruled out periodic orbits in the infinite-population limit, we now relax this assumption. As illustrated in Fig.~\ref{fig:Trajectories} (a), finite-size stochasticity transforms the damped oscillations of the deterministic dynamics into sustained macroscopic fluctuations. 
This phenomenon, known as resonant amplification \cite{mckane_predator-prey_2005, bladon_evolutionary_2010}, arises because intrinsic noise continuously excites the coupled resource–strategy modes and prevents the system from settling into its stable equilibrium. 

Unlike previous eco-evolutionary models \cite{weitz_oscillating_2016, tilman_evolutionary_2020}, where strategy update rules yield periodic orbits even in the deterministic limit, here persistent quasi-cycles emerge only as a finite-size effect, sustained entirely by noise.
To quantify the frequency of these noise-induced quasi-cycles, we employ the Linear Noise Approximation (LNA) \cite{gardiner_stochastic_2009, mckane_predator-prey_2005} to analytically derive the system's power spectrum (see SM).
We find that the spectrum of the resource exhibits a resonant peak at $f_{\mathrm{peak}}=\omega_\rho^*/2\pi$, with
$\omega_\rho^{*2}=-n+\sqrt{(n+\Delta)^2-n\tau^2}$. Because this frequency is governed by the noise-to-coupling ratio $n \propto  \nu K/N$, it depends on the relative timescale of strategic adaptation, $\nu$.
Fig.~\ref{fig:Trajectories} (b) plots $f_{\mathrm{peak}}$ as a function of $\nu$, demonstrating agreement between the theoretical LNA prediction and the frequencies extracted from numerical Gillespie simulations.

\medskip 
\subsection{Applications to real data of global fish landings.}
\begin{figure}
\centering
\includegraphics[width=1\linewidth]{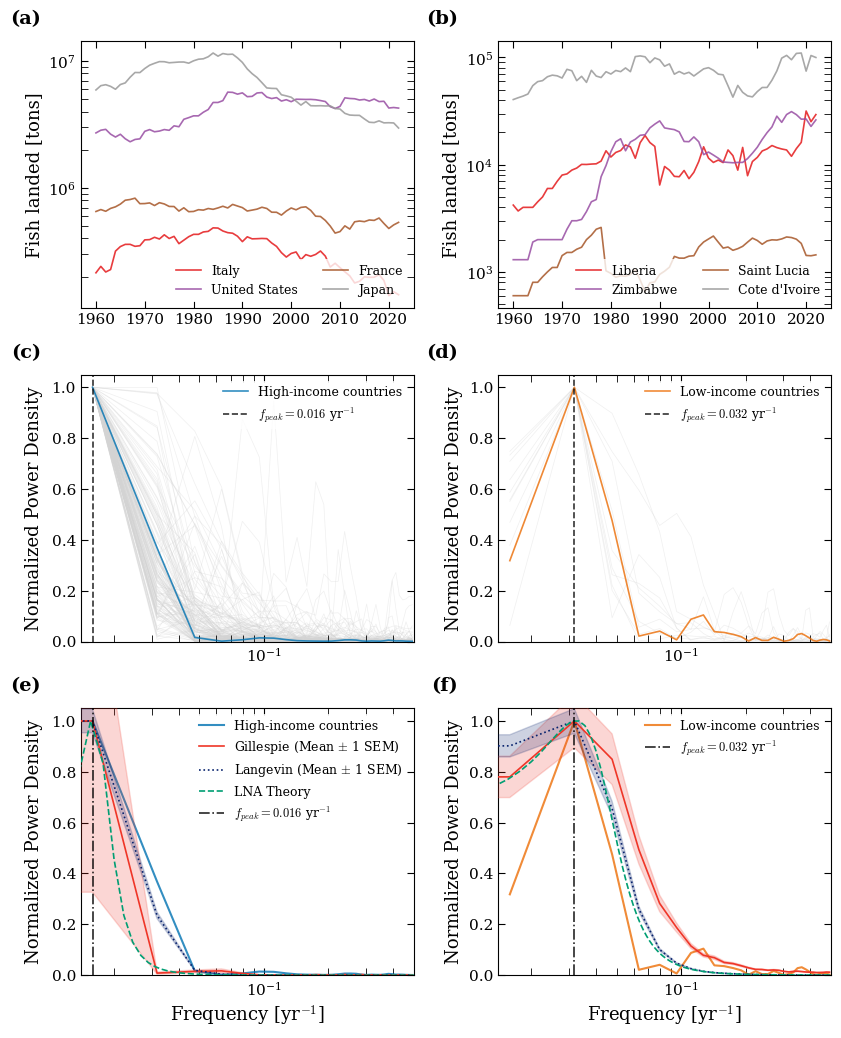}
\caption{
Temporal and spectral characterization of fish-landing dynamics. (a,b) Representative fish-landing time series (tons/year).
(c,d) Normalized power spectra  for all time series (light-gray lines) show dominant spectral peaks  at  $f = 0.016 \mathrm{yr}^{-1}$ (c) and $f = 0.032 \mathrm{yr}^{-1}$ (d). Solid curves denote the spectra for (c) high-income (teal) and (d) low-income (orange) countries. Income classifications follow World Bank definitions \cite{fao2026capture}.
(e,f) Spectra of empirical landings for (e) high-income (teal) and (f) low-income (orange) countries are compared to : analytical LNA (dashed green), Langevin integration (dotted blue, mean $\pm1$ s.e.m.) and Gillespie simulations (solid red, mean $\pm1$ s.e.m.). 
 The parameters adopted are (e) $e_{cc}=0.9$, $e_{dd}=1.8$, $\nu=0.24\,\mathrm{yr}^{-1}$, $K/N=100$ and (f) $e_{cc}=0.5$, $e_{dd}=1.5$, $\nu=0.38\,\mathrm{yr}^{-1}$, $K/N=100$.
Simulated trajectories are downsampled to match the annual temporal resolution of the empirical data, prior to spectral estimation.
}

\label{fig:Fig4}
\end{figure}

~ 
Surrogate-data tests and Kramers-Moyal diffusion estimates \cite{small_applied_2005, siegert_analysis_1998} applied to fish-landing time series from different countries \cite{fao2026capture} reveal that their fluctuations are not compatible with purely deterministic nonlinear dynamics, which points to intrinsic stochasticity (see SM).
To compare our model with the data, we interpret $N$ as the number of effective harvesting units (e.g.\ industrial fleets or vessels) and $K$ as the number of discrete biomass patches or quota units.
Reported landings are treated as a Poisson sample of the resource, i.e.\ proportional to resource availability.
Figure~\ref{fig:Fig4}(a-d) shows the landing time series and their corresponding spectra.
For 162 time series the spectral maximum lies at  $f = 0.016\,\mathrm{yr}^{-1}$, panel c of \ref{fig:Fig4}, which is the lowest resolved frequency ($1/63\,\mathrm{yr}^{-1}$) of the observation window. 
For 11 time series maximum lies at a higher frequency of $f = 0.032\,\mathrm{yr}^{-1}$, panel d of \ref{fig:Fig4} .
Fig.~\ref{fig:Fig4}(e-f) compares the aggregated empirical spectra for High- and Low-income countries against our theoretical framework (analytical LNA) alongside numerical Langevin and Gillespie simulations.
Both groups are reproduced by varying the noise ratio $n$ which depends on the extraction rates, the adaptation rate $\nu$ and the  ratio $K/N$.
High-income, industrialized fisheries (Fig.~\ref{fig:Fig4}(e)) combine intensive harvesting ($e_{cc}=e_{cd}=0.9$, $e_{dd}=e_{dc}=1.8$) with slower behavioral adaptation ($\nu=0.24\,\mathrm{yr}^{-1}$), and are run by few large operators (small $N$, $K/N=10000$). The system is then overdamped and the analytical spectrum decays monotonically from the lowest resolved frequency $f=0.016\,\mathrm{yr}^{-1}$, as observed.
Low-income, artisanal fisheries ($K/N=100$),  (Fig.~\ref{fig:Fig4}(f))  combine lighter extraction ($e_{cc}=e_{cd}=0.5$, $e_{dd}=e_{dc}=1.5$) with faster, local behavioral feedback ($\nu=0.38\,\mathrm{yr}^{-1}$ ). Here the LNA predicts an interior resonance at $f_{\mathrm{peak}}=0.032\,\mathrm{yr}^{-1}$, in agreement with the simulations and the empirical spectrum.

\medskip
\textbf{Conclusions.}
Real-world common-pool resources frequently exhibit complex oscillations amidst the tragedy of the commons \cite{kobeissi_tragedy_2024}. In this Letter we introduced a stochastic ECO-EVO modeling framework to study this dual behavior. Notice that  the strategy evolution of our model, in the quasi-stationary limit for slow resource dynamics, reduces to a replicator equation where resource feedback yields a Chicken (Hawk-Dove) game (see SM). By stabilizing a mixed strategy, this mechanism averts resource depletion despite a persistent local temptation to defect, providing a realistic bio-economic representation \cite{ito_complete_2024}. Furthermore, demographic noise resonantly amplifies the deterministic relaxation towards this state, sustaining macroscopic quasi-cycles. We found that the frequency of these fluctuations is controlled by the strategy-update timescale $\nu$ and the noise ratio $n$. By matching these parameters to the extraction and behavioral characteristics of high- and low-income countries, our model was able to successfully reproduce the frequency peaks found in global fish-landing data. Ultimately, our findings highlight the limits of purely deterministic models or of other  approaches that impose independent noise at the macroscopic scale, 
showing that intrinsic finite-size stochasticity and the timescales of human strategy adaptation must be treated as essential and intertwined drivers of socio-ecological sustainability.
\section{Acknowledgments}
We thank Sandro Azaele and Giacomo Frigerio for useful discussions. S. S. acknowledges financial support from the MUR - PNC (DD n. 1511
30-09-2022) Project No. PNC0000002, DigitAl lifelong pRevEntion (DARE). V.L. ackowledges NERC funding (NE/Y001184/1).
\section{Data Availability}
The code to reproduce all simulations and figures is available on GitHub at \url{https://github.com/SusannaCaruso/Resonant-
Amplification-of-Demographic-Noise-in-Stochastic-CPR-Dynamics}. The fish landing data used in this study are open-source and publicly available from the FAO \cite{fao2026capture}.

\bibliography{Reference}

@article{tilman_evolutionary_2020,
	title = {Evolutionary games with environmental feedbacks},
	volume = {11},
	copyright = {2020 The Author(s)},
	issn = {2041-1723},
	url = {https://www.nature.com/articles/s41467-020-14531-6},
	doi = {10.1038/s41467-020-14531-6},
	number = {1},
	urldate = {2026-03-19},
	journal = {Nature Communications},
	publisher = {Nature Publishing Group},
	author = {Tilman, Andrew R. and Plotkin, Joshua B. and Akçay, Erol},
	month = feb,
	year = {2020},
	pages = {915},
}

@article{weitz_oscillating_2016,
	title = {An oscillating tragedy of the commons in replicator dynamics with game-environment feedback},
	volume = {113},
	url = {https://www.pnas.org/doi/10.1073/pnas.1604096113},
	doi = {10.1073/pnas.1604096113},
	number = {47},
	urldate = {2026-03-19},
	journal = {Proceedings of the National Academy of Sciences},
	publisher = {Proceedings of the National Academy of Sciences},
	author = {Weitz, Joshua S. and Eksin, Ceyhun and Paarporn, Keith and Brown, Sam P. and Ratcliff, William C.},
	month = nov,
	year = {2016},
	pages = {E7518--E7525},
}

@article{saeedian_modelling_2024,
	title = {Modelling co-evolution of resource feedback and social network dynamics in human-environmental systems},
	volume = {26},
	issn = {1367-2630},
	url = {https://doi.org/10.1088/1367-2630/ad67fe},
	doi = {10.1088/1367-2630/ad67fe},
	number = {8},
	urldate = {2026-03-19},
	journal = {New Journal of Physics},
	publisher = {IOP Publishing},
	author = {Saeedian, Meghdad and Tu, Chengyi and Menegazzo, Fabio and D’Odorico, Paolo and Azaele, Sandro and Suweis, Samir},
	month = aug,
	year = {2024},
	pages = {083004},
}

@article{hardin_tragedy_1968,
	title = {The {Tragedy} of the {Commons}},
	volume = {162},
	url = {https://www.science.org/doi/10.1126/science.162.3859.1243},
	doi = {10.1126/science.162.3859.1243},
	number = {3859},
	urldate = {2026-03-22},
	journal = {Science},
	publisher = {American Association for the Advancement of Science},
	author = {Hardin, Garrett},
	month = dec,
	year = {1968},
	pages = {1243--1248},
}

@article{chavez_anchovies_2003,
	title = {From {Anchovies} to {Sardines} and {Back}: {Multidecadal} {Change} in the {Pacific} {Ocean}},
	volume = {299},
	shorttitle = {From {Anchovies} to {Sardines} and {Back}},
	url = {https://www.science.org/doi/10.1126/science.1075880},
	doi = {10.1126/science.1075880},
	number = {5604},
	urldate = {2026-03-22},
	journal = {Science},
	publisher = {American Association for the Advancement of Science},
	author = {Chavez, Francisco P. and Ryan, John and Lluch-Cota, Salvador E. and Ñiquen C., Miguel},
	month = jan,
	year = {2003},
	pages = {217--221},
}

@article{hutchings_collapse_2000,
	title = {Collapse and recovery of marine fishes},
	volume = {406},
	copyright = {2000 Macmillan Magazines Ltd.},
	issn = {1476-4687},
	url = {https://www.nature.com/articles/35022565},
	doi = {10.1038/35022565},
	number = {6798},
	urldate = {2026-03-22},
	journal = {Nature},
	publisher = {Nature Publishing Group},
	author = {Hutchings, Jeffrey A.},
	month = aug,
	year = {2000},
	pages = {882--885},
}

@article{myers_population_1995,
	address = {New York, N.Y.},
	title = {Population dynamics of exploited fish stocks at low population levels},
	volume = {269},
	issn = {0036-8075},
	doi = {10.1126/science.269.5227.1106},
	number = {5227},
	journal = {Science},
	author = {Myers, R. A. and Barrowman, N. J. and Hutchings, J. A. and Rosenberg, A. A.},
	month = aug,
	year = {1995},
	pages = {1106--1108},
}

@article{tu_emergence_2023,
	title = {The emergence of cooperation from shared goals in the governance of common-pool resources},
	volume = {6},
	copyright = {2022 The Author(s), under exclusive licence to Springer Nature Limited},
	issn = {2398-9629},
	url = {https://www.nature.com/articles/s41893-022-01008-1},
	doi = {10.1038/s41893-022-01008-1},
	number = {2},
	urldate = {2026-03-22},
	journal = {Nature Sustainability},
	publisher = {Nature Publishing Group},
	author = {Tu, Chengyi and D’Odorico, Paolo and Li, Zhe and Suweis, Samir},
	month = feb,
	year = {2023},
	pages = {139--147},
}

@article{mckane_predator-prey_2005,
	title = {Predator-{Prey} {Cycles} from {Resonant} {Amplification} of {Demographic} {Stochasticity}},
	volume = {94},
	url = {https://link.aps.org/doi/10.1103/PhysRevLett.94.218102},
	doi = {10.1103/PhysRevLett.94.218102},
	number = {21},
	urldate = {2026-03-22},
	journal = {Physical Review Letters},
	publisher = {American Physical Society},
	author = {McKane, A. J. and Newman, T. J.},
	month = jun,
	year = {2005},
	pages = {218102},
}

@article{gillespie_general_1976,
	title = {A general method for numerically simulating the stochastic time evolution of coupled chemical reactions},
	volume = {22},
	issn = {0021-9991},
	url = {https://www.sciencedirect.com/science/article/pii/0021999176900413},
	doi = {10.1016/0021-9991(76)90041-3},
	number = {4},
	urldate = {2026-03-22},
	journal = {Journal of Computational Physics},
	author = {Gillespie, Daniel T},
	month = dec,
	year = {1976},
	pages = {403--434},
}

@article{ito_complete_2024,
	title = {A complete classification of evolutionary games with environmental feedback},
	volume = {3},
	issn = {2752-6542},
	doi = {10.1093/pnasnexus/pgae455},
	number = {11},
	journal = {PNAS nexus},
	author = {Ito, Hiromu and Yamamichi, Masato},
	month = nov,
	year = {2024},
	pages = {pgae455},
}

@book{small_applied_2005,
	address = {Singapore},
	series = {World {Scientific} {Series} on {Nonlinear} {Science} {Series} {A} {Ser}},
	title = {Applied {Nonlinear} {Time} {Series} {Analysis}: {Applications} in {Physics}, {Physiology} and {Finance}},
	isbn = {978-981-256-117-6 978-981-256-777-2},
	shorttitle = {Applied {Nonlinear} {Time} {Series} {Analysis}},
	number = {v.52},
	publisher = {World Scientific Publishing Co Pte Ltd},
	author = {Small, Michael W.},
	year = {2005},
}

@article{siegert_analysis_1998,
	title = {Analysis of data sets of stochastic systems},
	volume = {243},
	issn = {0375-9601},
	url = {https://www.sciencedirect.com/science/article/pii/S0375960198002837},
	doi = {10.1016/S0375-9601(98)00283-7},
	number = {5},
	urldate = {2026-05-14},
	journal = {Physics Letters A},
	author = {Siegert, S and Friedrich, R and Peinke, J},
	month = jul,
	year = {1998},
	pages = {275--280},
}

@article{ostrom_challenge_2008,
	title = {The {Challenge} of {Common}-{Pool} {Resources}},
	url = {https://sci-hub.st/10.3200/ENVT.50.4.8-21},
	doi = {10.3200/ENVT.50.4.8-21},
	urldate = {2026-09-11},
	journal = {Environment: Science and Policy for Sustainable Development},
	author = {Ostrom, Elinor},
	year = {2008},
}

@article{sugihara_detecting_2012,
	title = {Detecting {Causality} in {Complex} {Ecosystems}},
	volume = {338},
	url = {https://www.science.org/doi/10.1126/science.1227079},
	doi = {10.1126/science.1227079},
	number = {6106},
	urldate = {2026-09-20},
	journal = {Science},
	publisher = {American Association for the Advancement of Science},
	author = {Sugihara, George and May, Robert and Ye, Hao and Hsieh, Chih-hao and Deyle, Ethan and Fogarty, Michael and Munch, Stephan},
	month = oct,
	year = {2012},
	pages = {496--500},
}

@article{gasull_effectiveness_2021,
	title = {Effectiveness of the {Bendixson}–{Dulac} theorem},
	volume = {305},
	issn = {0022-0396},
	url = {https://www.sciencedirect.com/science/article/pii/S0022039621006239},
	doi = {10.1016/j.jde.2021.10.011},
	urldate = {2026-09-30},
	journal = {Journal of Differential Equations},
	author = {Gasull, Armengol and Giacomini, Hector},
	month = dec,
	year = {2021},
	pages = {347--367},
}

@article{bendixson_sur_1901,
	title = {Sur les courbes définies par des équations différentielles},
	volume = {24},
	issn = {1871-2509},
	url = {https://doi.org/10.1007/BF02403068},
	doi = {10.1007/BF02403068},
	number = {1},
	urldate = {2026-09-30},
	journal = {Acta Mathematica},
	author = {Bendixson, Ivar},
	month = dec,
	year = {1901},
	pages = {1},
}

@article{bladon_evolutionary_2010,
	title = {Evolutionary dynamics, intrinsic noise, and cycles of cooperation},
	volume = {81},
	issn = {1539-3755},
	url = {https://research.manchester.ac.uk/en/publications/evolutionary-dynamics-intrinsic-noise-and-cycles-of-cooperation/},
	doi = {10.1103/PhysRevE.81.066122},
	number = {6},
	urldate = {2026-10-02},
	journal = {Physical Review E - Statistical, Nonlinear, and Soft Matter Physics},
	publisher = {American Physical Society},
	author = {Bladon, Alex J. and Galla, Tobias and McKane, Alan J.},
	month = jun,
	year = {2010},
	pages = {066122},
}

@book{kampen_stochastic_2011,
	title = {Stochastic {Processes} in {Physics} and {Chemistry}},
	isbn = {978-0-08-047536-3},
	publisher = {Elsevier},
	author = {Kampen, N. G. Van},
	month = aug,
	year = {2011},
}

@article{feeny_tragedy_1990,
	title = {The {Tragedy} of the {Commons}: {Twenty}-two years later},
	volume = {18},
	issn = {1572-9915},
	shorttitle = {The {Tragedy} of the {Commons}},
	url = {https://doi.org/10.1007/BF00889070},
	doi = {10.1007/BF00889070},
	number = {1},
	urldate = {2026-10-18},
	journal = {Human Ecology},
	author = {Feeny, David and Berkes, Fikret and McCay, Bonnie J. and Acheson, James M.},
	month = mar,
	year = {1990},
	pages = {1--19},
}

@incollection{ostrom_tragedy_2008,
	title = {Tragedy of the {Commons}},
	isbn = {978-1-349-95121-5},
	url = {https://link.springer.com/rwe/10.1057/978-1-349-95121-5_2047-1},
	urldate = {2026-10-18},
	booktitle = {The {New} {Palgrave} {Dictionary} of {Economics}},
	publisher = {Palgrave Macmillan, London},
	author = {Ostrom, Elinor},
	year = {2008},
	doi = {10.1057/978-1-349-95121-5_2047-1},
	pages = {1--5},
}

@article{kobeissi_tragedy_2024,
	title = {The tragedy of the commons: a {Mean}-{Field} {Game} approach to the reversal of travelling waves},
	volume = {37},
	issn = {0951-7715},
	shorttitle = {The tragedy of the commons},
	url = {https://doi.org/10.1088/1361-6544/ad7b97},
	doi = {10.1088/1361-6544/ad7b97},
	number = {11},
	urldate = {2026-10-18},
	journal = {Nonlinearity},
	publisher = {IOP Publishing},
	author = {Kobeissi, Ziad and Mazari-Fouquer, Idriss and Ruiz-Balet, Domènec},
	month = sep,
	year = {2024},
	pages = {115010},
}

@article{hauert_asymmetric_2019,
	title = {Asymmetric evolutionary games with environmental feedback},
	volume = {462},
	issn = {0022-5193},
	url = {https://www.sciencedirect.com/science/article/pii/S0022519318305721},
	doi = {10.1016/j.jtbi.2018.11.019},
	urldate = {2026-09-30},
	journal = {Journal of Theoretical Biology},
	publisher = {Elsevier},
	author = {Hauert, Christoph and Saade, Camille and McAvoy, Alex},
	month = feb,
	year = {2019},
	pages = {347--360},
}

@article{jonzen_theory_2002,
	title = {A {Theory} of {Stochastic} {Harvesting} in {Stochastic} {Environments}},
	volume = {159},
	url = {https://www.journals.uchicago.edu/doi/full/10.1086/339456},
	doi = {10.1086/339456},
	number = {5},
	urldate = {2026-10-19},
	journal = {The American Naturalist},
	publisher = {The University of Chicago Press},
	author = {Jonzén, Niclas and Ripa, Jörgen and Lundberg, Per},
	month = may,
	year = {2002},
	pages = {427--437},
}

@article{anderson_why_2008,
	title = {Why fishing magnifies fluctuations in fish abundance},
	volume = {452},
	copyright = {2008 Springer Nature Limited},
	issn = {1476-4687},
	url = {https://www.nature.com/articles/nature06851},
	doi = {10.1038/nature06851},
	number = {7189},
	urldate = {2026-10-19},
	journal = {Nature},
	publisher = {Nature Publishing Group},
	author = {Anderson, Christian N. K. and Hsieh, Chih-hao and Sandin, Stuart A. and Hewitt, Roger and Hollowed, Anne and Beddington, John and May, Robert M. and Sugihara, George},
	month = apr,
	year = {2008},
	pages = {835--839},
}

@book{gardiner_stochastic_2009,
	title = {Stochastic methods : a handbook for the natural and social sciences},
	isbn = {978-3-540-70712-7},
	shorttitle = {Stochastic methods},
	url = {http://archive.org/details/stochasticmethod0000gard},
	urldate = {2026-10-19},
	publisher = {Berlin : Springer},
	author = {Gardiner, C. W. (Crispin W. )},
	collaborator = {{Internet Archive}},
	year = {2009},
}

@article{zhao_stochastic_2026,
	title = {Stochastic ecoevolutionary dynamics under coupled behavioral and environmental feedback},
	volume = {520},
	issn = {0096-3003},
	url = {https://www.sciencedirect.com/science/article/pii/S0096300326000111},
	doi = {10.1016/j.amc.2026.129959},
	urldate = {2026-10-22},
	journal = {Applied Mathematics and Computation},
	author = {Zhao, Xiaoqian and Hu, Kaipeng and Tao, Yewei and Shi, Lei},
	month = jul,
	year = {2026},
	pages = {129959},
}

@misc{bodin_eco-evolutionary_2026,
	title = {Eco-evolutionary games in noisy environments},
	copyright = {© 2026, Posted by openRxiv. This pre-print is available under a Creative Commons License (Attribution-NonCommercial-NoDerivs 4.0 International), CC BY-NC-ND 4.0, as described at http://creativecommons.org/licenses/by-nc-nd/4.0/},
	url = {https://www.biorxiv.org/content/10.64898/2026.05.20.726658v1},
	doi = {10.64898/2026.05.20.726658},
	urldate = {2026-10-22},
	publisher = {bioRxiv},
	author = {Bodin, Fantine and Wang, Guocheng and Plotkin, Joshua B.},
	month = may,
	year = {2026},
	note = {ISSN: 2692-8205
Pages: 2026.05.20.726658
Section: New Results},
}

@dataset{fao2026capture,
author       =  {{{{Food and Agriculture Organization of the United Nations (FAO)}} and {{World Bank}}}},
title        = {Capture fishery production},
year         = {2026},
organization = {Our World in Data},
note         = {Processed by Our World in Data. Original data from \textit{World Development Indicators 129}. Archived on September 10, 2026},
urldate      = {2026-10-27},
url          = {https://archive.ourworldindata.org/20260910-011030/grapher/capture-fishery-production.html}
}

@article{yoshimoto_new_1996,
	title = {A new stochastic model for harvesting behavior with application to nonstationary forest growth and supply},
	volume = {26},
	issn = {0045-5067},
	url = {https://cdnsciencepub.com/doi/abs/10.1139/x26-222},
	doi = {10.1139/x26-222},
	number = {11},
	urldate = {2026-10-28},
	journal = {Canadian Journal of Forest Research},
	publisher = {NRC Research Press},
	author = {Yoshimoto, Atsushi},
	month = nov,
	year = {1996},
	pages = {1967--1972},
}

@article{tabi_species_2024,
	title = {Species interactions drive continuous assembly of freshwater communities in stochastic environments},
	volume = {14},
	copyright = {2024 The Author(s)},
	issn = {2045-2322},
	url = {https://www.nature.com/articles/s41598-024-72405-z},
	doi = {10.1038/s41598-024-72405-z},
	number = {1},
	urldate = {2026-10-28},
	journal = {Scientific Reports},
	publisher = {Nature Publishing Group},
	author = {Tabi, Andrea and Siqueira, Tadeu and Tonkin, Jonathan D.},
	month = sep,
	year = {2024},
	pages = {21747},
}

\newpage
\onecolumngrid

\newpage
\clearpage

\begin{center}
\textbf{\large Supplemental Material}\\[2pt]
\textbf{Resonant Amplification of Demographic Noise in Stochastic Common-Pool Resource Dynamics}
\end{center}

\setcounter{section}{0}
\setcounter{subsection}{0}
\setcounter{equation}{0}
\setcounter{figure}{0}
\renewcommand{\thesection}{S\arabic{section}}
\renewcommand{\theequation}{S\arabic{equation}}
\renewcommand{\thefigure}{S\arabic{figure}}
\renewcommand{\theHsection}{S\arabic{section}}
\renewcommand{\theHequation}{S\arabic{equation}}
\renewcommand{\theHfigure}{S\arabic{figure}}
\section{From the Microscopic Processes to the Langevin equation}
\subsection{Master Equation}

From the microscopic processes, we define the transition rates \(T(\mathbf{y} \mid \mathbf{z})\) representing the probability per unit time of transitioning from state \(\mathbf{z}\) to state \(\mathbf{y}\). Given the total population \(N = C + D\) and resource carrying capacity \(K\), the non-zero transition rates are:
\begin{equation}
\begin{aligned}
T(C, D, R+1 \mid C, D, R) &= b\,R\left(1-\frac{R}{K}\right), \\[4pt]
T(C, D, R-1 \mid C, D, R) &= h_c(C,D,R)\,C + h_d(C,D,R)\,D, \\[4pt]
T(C+1,D-1,R\,|\,C,D,R)&=\nu\,(1-R')^+\,\frac{D\,C}{N},\\[4pt]
T(C-1,D+1,R\,|\,C,D,R)&=\nu\,R'\,\frac{C\,D}{N},
\quad R'\equiv\frac{R}{K(1-e_{cc}/b)}\ge0\\ ,
\end{aligned}
\end{equation}
 with $(u)^+=\max(0,u)$. Here, $R_c  = K(1-e_{cc}/b)$ represents the effective carrying capacity of the resource under a fully cooperative population ($C=N, D=0$). Following the knowledge-feedback mechanism \cite{saeedian_modelling_2024}, agents assess resource abundance relative to this cooperative baseline: the propensity to defect scales with $R' = R/R_c$, whereas the scarcity-driven incentive to cooperate scales with $1-R'$. Because the carrying capacity is $K > R_c$, transient states with $R \in (R_c, K]$ correspond to $R' \in (1, (1-e_{cc}/b)^{-1}]$. In such states of abundance above the cooperative baseline, there is no perceived scarcity, and the operator $(1-R')^+$ sets the $d \to c$ transition rate to zero. Since $h_{c,d} \ge 0$ by construction, all transition rates are strictly non-negative across the entire state space and vanish at the physical boundaries $C,D \in \{0,N\}$ and $R \in \{0,K\}$.

The corresponding master equation governing the probability distribution \(P(C,D,R,t)\) reads:
\begin{equation}
\begin{aligned}
\frac{d}{dt} P(C,D,R,t) ={}& \Big[ (\mathcal{E}^r_- - 1) \, T(C,D,R+1 \mid C,D,R) \\
&+ (\mathcal{E}^r_+ - 1) \, T(C,D,R-1 \mid C,D,R) \\
&+ (\mathcal{E}^c_- \mathcal{E}^d_+ - 1) \, T(C+1,D-1,R \mid C,D,R) \\
&+ (\mathcal{E}^c_+ \mathcal{E}^d_- - 1) \, T(C-1,D+1,R \mid C,D,R) \Big] P(C,D,R,t),
\end{aligned}
\end{equation}
where the step operators act as \(\mathcal{E}^X_{\pm} f(X) = f(X \pm 1)\).

To obtain the continuum limit, we introduce the intensive variables:
\begin{equation}
x = \frac{C}{N}, \quad 1-x = \frac{D}{N}, \quad \rho = \frac{R}{K},
\end{equation}
with step increments \(\Delta x = \frac{1}{N}\) and \(\Delta \rho = \frac{1}{K}\). Under this scaling, the transition rates per unit volume/carrying capacity scale as:
\begin{equation}
\begin{aligned}
T(C,D,R+1 \mid C,D,R) &= K\, w_\rho^+(x, \rho) = K\, b\,\rho(1-\rho), \\[3pt]
T(C,D,R-1 \mid C,D,R) &= K\, w_\rho^-(x, \rho) = K\, \rho\, g(x), \\[3pt]
T(C+1,D-1,R \mid C,D,R) &= N\, w_x^+(x, \rho) = N\,\nu(1-\rho')^+\,x(1-x) \\[3pt]
T(C-1,D+1,R \mid C,D,R) &= N\, w_x^-(x, \rho) = N\, \nu \rho'\, x(1-x),
\end{aligned}
\end{equation}
where \(g(x) = e_{cc}x^2 + (e_{cd}+e_{dc})x(1-x) + e_{dd}(1-x)^2\) and \(\rho' = \frac{\rho}{1 - e_{cc}/b}\).

The discrete step operators translate into continuous shift operators:
\[
\begin{aligned}
&\mathcal{E}^r_{\pm} \to \epsilon^{ \pm \Delta\rho}, \qquad \epsilon^{ \pm \Delta\rho} f(\rho)= f(\rho \pm \Delta \rho)\\ &\mathcal{E}^c_{\pm}\mathcal{E}^d_{\mp} \to \epsilon^{\pm \Delta x} \qquad \epsilon^{\pm \Delta x} f(x) = f(x \pm \Delta x).
\end{aligned}
\]
Expanding the operators to second order in \(\Delta x = 1/N \ll 1\) and \(\Delta\rho = 1/K \ll 1\):
\begin{equation}
\begin{aligned}
\epsilon^{ \pm \Delta\rho} f(\rho) - 1 &= \pm \Delta\rho \frac{\partial}{\partial\rho} f(\rho) + \frac{\Delta\rho^2}{2} \frac{\partial^2}{\partial\rho^2}f(\rho) + \mathcal{O}(\Delta\rho^3 f(\rho)), \\[4pt]
\epsilon^{\pm \Delta x} f(x) - 1 &= \pm \Delta x \frac{\partial}{\partial x} f(x) + \frac{\Delta x^2}{2} \frac{\partial^2}{\partial x^2} f(x) + \mathcal{O}(\Delta x^3 f(x)).
\end{aligned}
\end{equation}

Substituting these expansions into the master equation yields:
\begin{equation}
\begin{aligned}
\frac{\partial}{\partial t} P(x,\rho,t) ={}& \left( -\frac{1}{K}\frac{\partial}{\partial\rho} + \frac{1}{2K^2}\frac{\partial^2}{\partial\rho^2} \right) \big[ K w_\rho^+(x,\rho) P \big] \\
&+ \left( +\frac{1}{K}\frac{\partial}{\partial\rho} + \frac{1}{2K^2}\frac{\partial^2}{\partial\rho^2} \right) \big[ K w_\rho^-(x,\rho) P \big] \\
&+ \left( -\frac{1}{N}\frac{\partial}{\partial x} + \frac{1}{2N^2}\frac{\partial^2}{\partial x^2} \right) \big[ N w_x^+(x,\rho) P \big] \\
&+ \left( +\frac{1}{N}\frac{\partial}{\partial x} + \frac{1}{2N^2}\frac{\partial^2}{\partial x^2} \right) \big[ N w_x^-(x,\rho) P \big].
\end{aligned}
\end{equation}

Simplifying the scale factors \(N\) and \(K\) and grouping by derivative orders leads directly to the two-dimensional Fokker--Planck equation:
\begin{equation}
\frac{\partial P}{\partial t} = -\frac{\partial}{\partial x} \big( A^x P \big) - \frac{\partial}{\partial \rho} \big( A^\rho P \big) + \frac{1}{2} \frac{\partial^2}{\partial x^2} \big( B^{xx} P \big) + \frac{1}{2} \frac{\partial^2}{\partial \rho^2} \big( B^{\rho\rho} P \big),
\end{equation}
where the drift coefficients \(A\) and diffusion coefficients \(B\) are:
\begin{equation}
\begin{aligned}
A^x(\rho,x) &= w_x^+ - w_x^- = \nu x(1-x)\big[(1-\rho')_+-\rho'\big], \\[4pt]
A^\rho(\rho,x) &= w_\rho^+ - w_\rho^- = b\rho(1-\rho) - \rho g(x), \\[4pt]
B^{xx}(\rho,x) &= \frac{1}{N} \big(w_x^+ + w_x^-\big) = \tfrac{\nu}{N}x(1-x)\big[(1-\rho')^++\rho'\big], \\[4pt]
B^{\rho\rho}(\rho,x) &= \frac{1}{K} \big(w_\rho^+ + w_\rho^-\big) = \frac{1}{K} \Big[ b\rho(1-\rho) + \rho g(x) \Big].
\end{aligned}
\end{equation}
Evaluating the positive-part operator $(1-\rho')^+$ explicitly in the two regions $\rho' \le 1$ (i.e., $\rho \le \rho_c \equiv 1-e_{cc}/b$) and $\rho' > 1$ ($\rho > \rho_c$), the strategy drift $A^x$ and unscaled diffusion factor $B_x(\rho,x) \equiv N B^{xx}(x,\rho)$ take the piecewise form
\begin{equation}
A^x(x,\rho) = \begin{cases}
\nu x(1-x)(1-2\rho') & \text{for } \rho' \le 1, \\
-\nu x(1-x)\rho' & \text{for } \rho' > 1,
\end{cases}
\label{eq:piecewise_A}
\end{equation}
\begin{equation}
B_x(\rho,x)  = \begin{cases}
\nu x(1-x) & \text{for } \rho' \le 1, \\
\nu x(1-x)\rho' & \text{for } \rho' > 1.
\end{cases}
\label{eq:piecewise_B}
\end{equation}

\subsection{Derivation of the Coupled Langevin System (It\^o Interpretation)}

To derive the continuous stochastic dynamics corresponding to the multivariate Fokker--Planck equation, we consider the state vector $\mathbf{z} = (x, \rho)^\top$. The general multivariate Fokker--Planck equation takes the form:
\begin{equation}
\frac{\partial P(\mathbf{z}, t)}{\partial t} = -\sum_{i \in \{x, \rho\}} \frac{\partial}{\partial z_i} \Big[ A^i(\mathbf{z}) P(\mathbf{z}, t) \Big] + \frac{1}{2} \sum_{i, j \in \{x, \rho\}} \frac{\partial^2}{\partial z_i \partial z_j} \Big[ B^{ij}(\mathbf{z}) P(\mathbf{z}, t) \Big],
\end{equation}
where $A^i(\mathbf{z})$ denotes the deterministic drift vector and $B^{ij}(\mathbf{z})$ is the symmetric diffusion tensor. Because transitions in the microscopic coordinate $x$ and the resource density $\rho$ proceed via statistically independent jump events, the off-diagonal cross-diffusion coefficients vanish identically ($B^{x\rho} = B^{\rho x} = 0$). The diffusion matrix is therefore diagonal:
\begin{equation}
\mathbf{B}(x,\rho) = 
\begin{pmatrix}
B^{xx}(x,\rho) & 0 \\
0 & B^{\rho\rho}(x,\rho)
\end{pmatrix}
=
\begin{pmatrix}
\frac{\nu}{N} x(1-x)\big[(1-\rho')^+ + \rho'\big] & 0 \\[6pt]
0 & \frac{\rho}{K} \big[ b(1-\rho) + g(x) \big]
\end{pmatrix}.
\end{equation}

Under the It\^o interpretation, the Fokker--Planck equation maps directly to a system of stochastic differential equations \cite{kampen_stochastic_2011,gardiner_stochastic_2009}. The noise amplitude matrix $\boldsymbol{\sigma}(\mathbf{z})$ satisfies $\mathbf{B}(\mathbf{z}) = \boldsymbol{\sigma}(\mathbf{z}) \boldsymbol{\sigma}^\top(\mathbf{z})$. Choosing the diagonal representation $\sigma_{ii}(\mathbf{z}) = \sqrt{B^{ii}(\mathbf{z})}$, the system reads in It\^o Langevin form:
\begin{equation}
\begin{aligned}
\frac{dx}{dt} &= A^x(x,\rho) + \sqrt{B^{xx}(x,\rho)} \, \eta_x(t), \\[6pt]
\frac{d\rho}{dt} &= A^\rho(x,\rho) + \sqrt{B^{\rho\rho}(x,\rho)} \, \eta_\rho(t),
\end{aligned}
\end{equation}
where $\eta_x(t)$ and $\eta_\rho(t)$ are mutually independent standard Gaussian white noise processes with zero mean and covariance $\langle \eta_i(t) \eta_j(t') \rangle = \delta_{ij}\delta(t - t')$. 

Substituting the explicit drift and diffusion expressions yields the general coupled Langevin equations valid across the entire state space $\Omega = [0,1]^2$:
\begin{equation}
\begin{aligned}
\frac{dx}{dt} &= \nu x(1-x)\big[(1-\rho')^+-\rho'\big] + \sqrt{\frac{\nu x(1-x)\big[(1-\rho')_++\rho'\big]}{N}} \, \eta_x(t), \\[8pt]
\frac{d\rho}{dt} &= \rho\big[b(1-\rho) - g(x)\big] + \sqrt{\frac{\rho\big[b(1-\rho) + g(x)\big]}{K}} \, \eta_\rho(t).
\end{aligned}
\label{eq:general_langevin_SM}
\end{equation}
In the region $\rho \le \rho_c \equiv 1-e_{cc}/b$ (i.e., $\rho' \le 1$), Eq.~(\ref{eq:general_langevin_SM}) reduces to Eq.~(2) of the main text, or equivalently in It\^o differential notation:
\begin{equation}
\begin{aligned}
dx &= \nu x(1-x)(1-2\rho') \, dt + \sqrt{\frac{\nu x(1-x)}{N}} \, dW_t^x, \\[8pt]
d\rho &= \rho\big[ b(1-\rho) - g(x) \big] \, dt + \sqrt{\frac{\rho\big[b(1-\rho) + g(x)\big]}{K}} \, dW_t^\rho,
\end{aligned}
\end{equation}
where $W_t^x$ and $W_t^\rho$ denote independent standard Wiener processes.

\subsection{Domain of Validity of Eq.~(2) of the Main Text}
In the main text, Eq.~(2) is presented in its simplified form corresponding to the region $\rho \le \rho_c  = 1 - e_{cc}/b$ (or equivalently $\rho' \le 1$), where the positive-part operator simply reduces to $(1-\rho')^+ = 1-\rho'$. Here we show why the system resides in this region and why the cutoff at $\rho = \rho_c$ does not alter any of our analytical results.

First, consider how the resource $\rho$ evolves across the state space. Because cooperators extract the resource at the lowest rate ($e_{cc} \le e_{cd}, e_{dc}$ and $e_{cc} < b < e_{dd}$), the total harvest pressure $g(x)$ is minimized when the population consists solely of cooperators ($x=1$), where $g(1) = e_{cc}$. Whenever defectors are present ($x < 1$), the harvest pressure is higher:
\begin{equation}
g(x) = e_{cc}x^2 + (e_{cd}+e_{dc})x(1-x) + e_{dd}(1-x)^2 \ge e_{cc}\big[x^2 + 2x(1-x) + (1-x)^2\big] = e_{cc}.
\end{equation}
Using the definition $\rho_c = 1 - e_{cc}/b$ (so that $e_{cc} = b(1-\rho_c)$), we can bound the deterministic rate of change of the resource as
\begin{equation}
\dot\rho = \rho\,\big[b(1-\rho)-g(x)\big] \le \rho\,\big[b(1-\rho)-e_{cc}\big] = b\rho(\rho_c-\rho),
\end{equation}
with strict inequality whenever $x < 1$. This simple bound has two direct physical consequences:
\begin{itemize}
\item[(i)] \textbf{Trajectories are  confined to $\rho \le \rho_c$:} Whenever the resource exceeds the cooperative carrying capacity ($\rho > \rho_c$), the right-hand side of the inequality above is negative, meaning $\dot\rho < 0$ regardless of the strategy composition $x$. Even along the threshold line $\rho = \rho_c$, the resource decreases ($\dot\rho < 0$) for any $x < 1$. Therefore, any deterministic trajectory that starts inside the domain $\Omega_c = [0, \rho_c] \times [0,1]$ can never leave it. Conversely, if the system is initialized in a state of abundant resource with $\rho(0) > \rho_c$ and $x(0) < 1$, the resource steadily declines and because abundance promotes defection ($\dot x < 0$), harvesting intensifies during the transient, pulling the system into the region $\rho \le \rho_c$ in finite time.
\item[(ii)] \textbf{All equilibria and local stability properties are unaffected:} Within $\Omega_c$, the general Langevin system in Eq.~(\ref{eq:general_langevin_SM}) is identical to Eq.~(2) of the main text. All physically admissible fixed points lie inside $\Omega_c$. In particular, the coexistence equilibrium $(\rho^*, x^*)$ sits right in the middle of this domain at $\rho^* = \rho_c / 2$ (so that $\rho'^* = 1/2$). Consequently, local expansions around the interior equilibrium, including the Jacobian stability analysis and the Linear Noise Approximation (LNA), depend solely on the smooth dynamics within $\rho < \rho_c$. 
\item[(iii)] \textbf{Stochastic fluctuations into $\rho > \rho_c$ are exponentially rare at large $K$:} In a finite system, demographic noise causes $\rho$ to fluctuate around the interior equilibrium $\rho^* = \rho_c / 2$ with a typical amplitude that shrinks as $1/\sqrt{K}$. Because the threshold $\rho = \rho_c$ lies at a fixed distance $\Delta\rho = \rho_c/2$ from the equilibrium, crossing it requires a fluctuation many standard deviations above the mean. The probability of such a large excursion decays exponentially with $K$ \cite{kampen_stochastic_2011,gardiner_stochastic_2009}, meaning that after the initial transient the system remains inside $\rho \le \rho_c$ virtually all the time.
\end{itemize}

\label{Sec:Validity}
\section{Linear Stability Analysis and Jacobian}

Within the domain $\rho \le \rho_c$ ($\rho' \le 1$), the Jacobian of the deterministic system is given by
\begin{equation}
J(\rho,x) =
\begin{pmatrix}
\dfrac{\partial \dot \rho}{\partial \rho} & \dfrac{\partial \dot \rho}{\partial x} \\[3mm]
\dfrac{\partial \dot x}{\partial \rho} & \dfrac{\partial \dot x}{\partial x}
\end{pmatrix},
\end{equation}
with entries
\[
\begin{aligned}
\dfrac{\partial \dot \rho}{\partial \rho} &= b(1-2\rho) - g(x),\\
\dfrac{\partial \dot \rho}{\partial x} &= -\rho\,g'(x) = -\rho\Big[ 2 e_{cc} x + (e_{cd}+e_{dc})(1-2x) - 2 e_{dd} (1-x) \Big],\\[1mm]
\dfrac{\partial \dot x}{\partial \rho} &= - \dfrac{2 \nu x(1-x)}{1-e_{cc}/b},\\[1mm]
\dfrac{\partial \dot x}{\partial x} &= \nu (1-2x)(1-2\rho'),
\end{aligned}
\]
where $g(x) = e_{cc} x^2 + (e_{cd}+e_{dc})\, x (1-x) + e_{dd} (1-x)^2$.

Evaluated at the interior equilibrium $(\rho^*, x^*)$, where $b(1-\rho^*) = g(x^*)$ and $\rho'^* = 1/2$ (so that $J_{xx} = 0$), the Jacobian reduces to
\begin{equation}
J =
\begin{pmatrix}
-b\rho^* & -\rho^* g'(x^*)\\[1mm]
-a & 0
\end{pmatrix},
\qquad
a = \frac{2\nu x^*(1-x^*)}{1-e_{cc}/b} > 0.
\end{equation}
The trace and determinant of $J$ are therefore
\begin{equation}
\tau \equiv \mathrm{Tr}\,J = -b\rho^* < 0, \qquad
\Delta \equiv \det J = -a\,\rho^*g'(x^*) > 0,
\end{equation}
where $\Delta > 0$ follows from $g'(x^*) < 0$ (since increasing the fraction of cooperators reduces the overall harvest pressure). Because $\tau < 0$ and $\Delta > 0$, the interior fixed point is always locally asymptotically stable. Its eigenvalues are given by
\begin{equation}
\lambda_{\pm} = \frac{\tau}{2} \pm i\Omega, \qquad \Omega = \sqrt{\Delta - \frac{\tau^2}{4}}.
\end{equation}
Accordingly, the interior equilibrium is a stable node if $\tau^2 > 4\Delta$  and a stable focus/spiral if $\tau^2 < 4\Delta$ (complex conjugate eigenvalues with negative real part). Since $\tau\neq0$ at the interior fixed point, the linearized system never has a center and never undergoes a Hopf bifurcation, oscillations can only be sustained by noise (quasi-cycles), see also Sec.~\ref{sec:cycles-exclusion}.

\section{Exclusion of periodic orbits}
\label{sec:cycles-exclusion}
Having classified the local stability of the interior fixed point, we now prove that the deterministic (drift) dynamics of system (2) in the main text cannot admit limit cycles or closed periodic orbits. 
As shown in Sec \ref{Sec:Validity}, since $\dot\rho < 0$ everywhere in $\rho > \rho_c$, any hypothetical periodic orbit must lie  within the region $\rho \le \rho_c$.
In the open, simply connected interior $\mathcal{D} = (0, 1) \times (0, \rho_c) \subset \mathbb{R}^2$ (and more generally on $(0,1)^2$), the deterministic drift vector field $\mathbf{F}=(F(\rho,x),G(\rho,x))$ is defined by the planar system:
\begin{equation}
\begin{cases}
    \dot{x} = F(\rho,x) = \nu x(1-x)(1-2\rho'), \\[6pt]
    \dot{\rho} = G(\rho,x) = \rho \big[ b(1-\rho) - g(x) \big].
\end{cases}
\end{equation}
To rule out the existence of periodic solutions, we employ the Bendixson--Dulac theorem \cite{bendixson_sur_1901,gasull_effectiveness_2021} using the Dulac multiplier:
\begin{equation}
    \phi(\rho,x) = \frac{1}{\rho x (1-x)}.
\end{equation}
The function $\phi(x, \rho)$ is of class $\mathcal{C}^{\infty}$, strictly positive, and non-vanishing everywhere in $\mathcal{D}$.

According to the Bendixson-Dulac criterion, if the weighted divergence:
\begin{equation}
    \nabla \cdot (\phi \mathbf{F}) = \frac{\partial (\phi F)}{\partial x} + \frac{\partial (\phi G)}{\partial \rho}
\end{equation}
does not change sign and is not identically zero on any open subset of $\mathcal{D}$, then the system contains no closed periodic orbits entirely contained in $\mathcal{D}$. 

Evaluating each term explicitly:
\begin{align}
    \frac{\partial}{\partial x} \big( \phi F \big) &= \frac{\partial}{\partial x} \left[ \frac{\nu(1-2\rho')}{\rho} \right] = 0, 
    \label{eq:phiF}\\[8pt]
    \frac{\partial}{\partial \rho} \big( \phi G \big) &= \frac{\partial}{\partial \rho} \left[ \frac{b(1-\rho) - g(x)}{x(1-x)} \right] = -\frac{b}{x(1-x)}.
\end{align}
Therefore, the divergence simplifies to:
\begin{equation}
    \nabla \cdot (\phi \mathbf{F}) = -\frac{b}{x(1-x)}.
\end{equation}
Given that $b > 0$ and $x(1-x) > 0$ for all $x \in (0, 1)$, it follows that $\nabla \cdot (\phi \mathbf{F}) < 0$ strictly across the entire domain $\mathcal{D}$. Consequently, by the Bendixson--Dulac theorem, the deterministic system does not admit any periodic orbits or limit cycles.
Remarkably, this holds even if in (\ref{eq:phiF}) $(1-2\rho')$ is replaced by the full expression $[(1-\rho')_+ - \rho']$ on $(0,1)^2$.
\newpage

\section{Linear-Noise Approximation: Analytical Power Spectrum}
\label{sec:LNA}
\subsection{Linearization around the Interior Fixed Point}
To quantify the stochastic quasi-cycles analytically, we apply the Linear Noise Approximation (LNA) \cite{kampen_stochastic_2011,gardiner_stochastic_2009} around the interior equilibrium $(\rho^*, x^*)$. This local expansion requires that the model parameters admit a feasible interior equilibrium $x^* \in (0,1)$, which sits at $\rho^* = (1 - e_{cc}/b)/2 = \rho_c/2$ (i.e., $\rho'^* = 1/2$, hence $\rho' < 1$), and the system sizes $K, N \gg 1$ are sufficiently large that fluctuations around $(\rho^*, x^*)$ remain small, of order $\mathcal{O}(K^{-1/2}, N^{-1/2})$.

Let $\boldsymbol z = (\delta\rho, \delta x)^\top$ denote the fluctuations around the interior fixed point $(\rho^*, x^*)$, where $g(x^*) = b(1 - \rho^*)$. To leading order in $K^{-1/2}$ and $N^{-1/2}$, the linearized Langevin dynamics reads
\begin{equation}
\dot{\boldsymbol z} = J\boldsymbol z + \boldsymbol\xi(t), \qquad
\langle \boldsymbol\xi(t)\boldsymbol\xi(t')^\top \rangle = B\,\delta(t-t'),
\end{equation}
where $J$ is the Jacobian evaluated at $(\rho^*, x^*)$, and the noise covariance (diffusion) matrix $B = \mathbf{B}(\rho^*, x^*)$ accounting for demographic noise in both variables is
\begin{equation}
B = \begin{pmatrix} B_{\rho} & 0 \\ 0 & B_{x} \end{pmatrix}
= \begin{pmatrix} \dfrac{2b\rho^*(1-\rho^*)}{K} & 0 \\[3mm] 0 & \dfrac{\nu x^*(1-x^*)}{N} \end{pmatrix},
\end{equation}
where we used $b\rho^*(1-\rho^*) + \rho^*g(x^*) = 2b\rho^*(1-\rho^*)$.

\subsection{Spectral Matrix and Resonant Peak Frequency}
To compute the power spectral density matrix, we take the Fourier transform $\tilde{\boldsymbol z}(\omega) = \int_{-\infty}^{\infty} \boldsymbol z(t)e^{-i\omega t}dt$.
The spectral matrix is given by
\begin{equation}
S(\omega) = M^{-1}B\,(M^{-1})^\dagger, \qquad M = i\omega I - J,
\end{equation}
with
\begin{equation}
M^{-1} = \frac{1}{\det M}
\begin{pmatrix} i\omega - J_{xx} & J_{\rho x} \\ J_{x\rho} & i\omega - J_{\rho\rho} \end{pmatrix},
\qquad
\det M = (\Delta - \omega^2) - i\omega\tau,
\end{equation}
such that
\begin{equation}
|\det M|^2 = (\Delta - \omega^2)^2 + \omega^2\tau^2 \equiv \mathcal D(\omega).
\end{equation}
Using $J_{xx} = 0$ and $J_{\rho\rho} = \tau$, the elements of the spectral matrix evaluate to:
\begin{align}
S_{\rho\rho}(\omega) &= \frac{\omega^2 B_\rho + J_{\rho x}^2 B_x}{\mathcal D(\omega)}, \label{eq:S_rhorho}\\
S_{xx}(\omega) &= \frac{(\omega^2 + \tau^2)B_x + J_{x\rho}^2 B_\rho}{\mathcal D(\omega)}, \\
S_{\rho x}(\omega) &= \frac{i\omega J_{x\rho} B_\rho - (i\omega + \tau)J_{\rho x} B_x}{\mathcal D(\omega)} = S_{x\rho}^*(\omega).
\end{align}

Because $S_{\rho\rho}(\omega)$ is an even function of $\omega$, we find its local extrema for $\omega > 0$ by introducing the variable $u = \omega^2 \ge 0$. In terms of $u$, Eq.~\eqref{eq:S_rhorho} becomes
\begin{equation}
S_{\rho\rho}(u) = \frac{u B_\rho + J_{\rho x}^2 B_x}{(\Delta - u)^2 + u\tau^2}.
\end{equation}
Setting $c \equiv \tau^2 - 2\Delta$ and defining the noise-to-coupling ratio $n \equiv \frac{J_{\rho x}^2 B_x}{B_\rho}$, this expression simplifies to
\begin{equation}
S_{\rho\rho}(u) = B_\rho \frac{u + n}{u^2 + cu + \Delta^2}.
\end{equation}
The stationary points satisfy $\frac{dS_{\rho\rho}}{du} = 0$, which yields the quadratic equation
\begin{equation}
\label{eq:u}
u^2 + 2nu - (\Delta^2 - nc) = 0.
\end{equation}
Using $\Delta^2 - nc = \Delta^2 - n(\tau^2 - 2\Delta) = (n + \Delta)^2 - n\tau^2$, the unique physical root is
\begin{equation}
\omega^{*2}_{\rho} = u^* = -n + \sqrt{(n + \Delta)^2 - n\tau^2}.
\label{eq:omega_star_sq}
\end{equation}
Analogously, the power spectrum of the strategies, $S_{xx}(\omega)$, peaks at
\begin{equation}
    \omega^{*2}_{x} = -\tilde n + \sqrt{(\tilde n + \Delta)^2 - \tilde n\tau^2},
\end{equation}
where $\tilde n = \tau^2 + \frac{J^2_{x\rho}B_\rho}{B_x}$.
A peak at $\omega_\rho>0$ exists iff
\begin{equation}
 \Delta^2>n( \tau^2-2\Delta);
\label{eq:peakexist}
\end{equation}
otherwise $S$ decreases monotonically from $\omega_{\rho}=0$.

\subsection{Comparison with the Deterministic Frequency $\Omega=|\mathrm{Im}\,\lambda|$}
Substituting $u = \Omega^2 \equiv \Delta - \tau^2/4$ into Eq.~\eqref{eq:u} with $c = \tau^2 - 2\Delta$ shows that the LNA peak frequency $f_{\mathrm{peak}} = \omega_\rho^*/(2\pi)$ coincides with the deterministic damped oscillation frequency $f_0 = \Omega/(2\pi)$ at the critical noise ratio
\begin{equation}
    n_c = \Delta - \frac{\tau^2}{8} = \Omega^2 + \frac{\tau^2}{8}.
\end{equation}
In the presence of demographic noise, the observed resonant frequency $\omega^*_{\rho}$ is shifted relative to the deterministic frequency $\Omega$ depending on $\tau$ and the relative intensity $n$ of the two noise sources:
\begin{itemize}
    \item \textbf{Resource-dominated noise ($n \to 0$, $n < n_c$, i.e., $J_{\rho x}^2 B_{x} \ll B_{\rho}$):} Taking $n \to 0$ in Eq.~\eqref{eq:omega_star_sq} gives
    \begin{equation}
    \omega^*_\rho \to \sqrt{\Delta} = \sqrt{\Omega^2 + \frac{\tau^2}{4}} > \Omega.
    \end{equation}
    In this limit, demographic noise shifts the spectral peak to a higher frequency than the deterministic frequency $\Omega$.
    
    \item \textbf{Strategy-dominated noise ($n \to \infty$, $n > n_c$, i.e., $B_\rho \ll J_{\rho x}^2 B_x$):} Expanding the square root in Eq.~\eqref{eq:omega_star_sq} for large $n$, $\sqrt{(n+\Delta)^2 - n\tau^2} = n + \Delta - \frac{\tau^2}{2} + \mathcal O(n^{-1})$, yields
    \begin{equation}
    \omega_{\rho}^* \to \sqrt{\Delta - \frac{\tau^2}{2}} = \sqrt{\Omega^2 - \frac{\tau^2}{4}} < \Omega.
    \end{equation}
    Here, demographic noise shifts the resonant frequency below the deterministic value (and suppresses the peak entirely if $\Delta \le \tau^2/2$).
\end{itemize}

\subsection{Closed Form in Terms of the Microscopic Parameters}
For the symmetric cross-interaction choice used in the main text, $e_{cd}+e_{dc}=e_{cc}+e_{dd}$, the harvest function simplifies to the linear form $g(x)=e_{cc}x+e_{dd}(1-x)$, with constant derivative $g'(x)=e_{cc}-e_{dd}<0$. The interior fixed point is given explicitly by
\begin{equation}
\rho^*=\frac{b-e_{cc}}{2b},\qquad x^*=\frac{\chi}{2},\qquad
\chi\equiv\frac{2e_{dd}-e_{cc}-b}{e_{dd}-e_{cc}}\in(0,2),
\end{equation}
so that $x^*(1-x^*)=\chi(b-e_{cc})/[4(e_{dd}-e_{cc})]$.
The Jacobian trace and determinant then read
\begin{equation}
\tau=-\frac{b-e_{cc}}{2},\qquad
\Delta=\nu x^*(1-x^*)(e_{dd}-e_{cc})=\frac{\nu\chi(b-e_{cc})}{4},
\end{equation}
while the deterministic frequency and resource noise intensity become
\begin{equation}
\Omega=\frac{\sqrt{b-e_{cc}}}{4}\sqrt{4\nu\chi-(b-e_{cc})},\qquad
B_{\rho}=\frac{b^2-e_{cc}^2}{2bK}.
\end{equation}
Consequently, the noise-to-coupling ratios take the compact forms
\begin{equation}
n=\frac{K}{N}\,\gamma\,\Delta,\qquad
\tilde n=\tau^2+\frac{N}{K}\frac{\Delta}{\gamma},\qquad \text{with} \qquad
\gamma\equiv\frac{(b-e_{cc})(e_{dd}-e_{cc})}{2b(b+e_{cc})}.
\end{equation}
Thus, $K/N\gg 1$ corresponds to strategy-dominated demographic noise, whereas $K/N\ll 1$ corresponds to resource-dominated demographic noise.

\newpage

\section{Time series analysis}
To assess the presence of stochastic dynamics in the data obtained from \citep{fao2026capture}, we performed a preliminary analysis of the time series using two complementary techniques: surrogate-data testing and the empirical estimation of the Kramers-Moyal diffusion coefficient.

The dataset consists of 229 time series spanning 63 years. These include both country-level estimates and aggregated classifications provided by the FAO and World Bank, namely four income groups (high-income, upper-middle-income, lower-middle-income, and low-income countries) and seven World Bank geographical regions.

Prior to the analysis, incomplete and non-informative time series were removed. Specifically, we discarded all series containing more than \(30\%\) zero entries (37 time series) as well as one constant series.

To apply the surrogate-data testing procedure, we first removed the long-term trend associated with the global increase in fishing activity observed over recent decades. To this end, the original series were transformed into log-returns. Following this transformation, additional time series containing more than \(20\%\) missing values (\(\mathrm{NaN}\)) were discarded. The remaining series were subsequently standardized, yielding a final dataset composed of 188 time series.

The surrogate-data analysis tests whether the observed fluctuations are compatible with a stochastic process or instead reflect non linear deterministic systems. In particular, we tested the null hypothesis that the data were generated by a monotonic nonlinear transformation of linearly filtered noise.

As test statistic, we employed the nonlinear prediction error (NLPE), which quantifies how accurately neighboring points in the reconstructed phase space predict the future evolution of a trajectory. The surrogate-testing procedure consists in comparing the value of the test statistic computed from the empirical data with that obtained from surrogate series preserving both the autocorrelation structure (equivalently, the power spectrum) and the probability distribution of the original data.
For each empirical time series, we generated 100 surrogate realizations. The nonlinear prediction error was computed using an embedding dimension \(m=3\) \citep{small_applied_2005}, delay \(t=1\), and \(k=3\) nearest neighbors. 

Under these parameters, only 2 out of the 188 time series exhibited a test statistic significantly different from that of the corresponding surrogate ensemble. In contrast, the remaining 186 series showed no statistically significant deviation from the surrogate data (median \(p\)-value: \(p = 0.465\)). To corroborate this result, we estimated the second Kramers--Moyal coefficient (diffusion coefficient $D^{(2)}$) directly from the standardized time series following \citep{siegert_analysis_1998}. Consistent with the surrogate test, we found a strictly positive diffusion coefficient across the dataset, with a median value $D^{(2)} = 0.2956$ and a mean value $\langle D^{(2)} \rangle = 0.3989$.
Although both methods are subject to finite-sample effects due to the 63-year observation window, together they provide empirical support for modeling fishery fluctuations as an intrinsically stochastic process.
Beyond confirming their stochastic nature, spectral analysis of the 188 valid fish-landing time series reveals that the majority  cluster into two main characteristic frequency groups. Specifically, 162 out of 188 series exhibit a dominant spectral peak at the lowest resolvable frequency of the 63-year observation window, $f = 1/63\,\mathrm{yr}^{-1} \approx 0.016\,\mathrm{yr}^{-1}$ ($0.01587\,\mathrm{yr}^{-1}$), corresponding to a monotonically decaying (overdamped) spectrum. A second coherent group, comprising 11 out of 188 series (including the low-income country aggregate), instead displays an interior resonant peak at $f = 2/63\,\mathrm{yr}^{-1} \approx 0.032\,\mathrm{yr}^{-1}$ ($0.03175\,\mathrm{yr}^{-1}$). The emergence of the two primary spectral groups points to distinct underlying eco-evolutionary feedbacks, which we link in the main text to differences in harvesting pressure, strategy-adaptation timescales and noise ratio.
\section{Mapping to the standard replicator equation}
Usually, eco-evolutionary models with environmental feedback explicitly incorporate strategy dynamics into a standard replicator equation by assuming a payoff matrix, or a linear combination, that varies according to the resource dynamics. Crucially, in our framework, the payoff matrix is not fixed beforehand. Instead, we demonstrate that by considering the mean-field system, i.e., assuming zero noise in the Langevin system (2) of the main text and analyzing the local behavior around the interior fixed point, we can directly map our model to a standard replicator equation.

Because the strategy dynamics are explicitly coupled to the resource, such a mapping holds in the quasi-stationary regime where the resource relaxes to its non-trivial equilibrium $\rho^*(x) = 1 - g(x)/b$ for a given strategy composition $x$.

We begin by recalling the classical formulation of replicator dynamics for two interacting strategies.
Consider a well-mixed population in which individuals engage in pairwise interactions described by the payoff matrix
\begin{equation}
\Pi =
\begin{pmatrix}
R & S \\[2mm]
T & P
\end{pmatrix}.
\end{equation}
Let \(x\in[0,1]\) denote the fraction of individuals adopting the first strategy.
The expected payoffs of the two strategies are
\begin{equation}
\pi_1(x) = R x + S (1-x), 
\qquad
\pi_2(x) = T x + P (1-x),
\end{equation}
and the mean population payoff is
\begin{equation}
\bar{\pi}(x) = x \pi_1(x) + (1-x)\pi_2(x).
\end{equation}
The replicator equation governing the evolution of \(x\) is then given by
\begin{equation}
\frac{dx}{dt}
= x\bigl(\pi_1(x)-\bar{\pi}(x)\bigr)
= x(1-x)\bigl(\pi_1(x)-\pi_2(x)\bigr).
\end{equation}
Substituting the explicit expressions for the payoffs yields
\begin{equation}
\frac{dx}{dt}
= x(1-x)\bigl(Ax+B\bigr),
\label{eq:standard_replicator}
\end{equation}
where
\[
A = R-S-T+P,
\qquad
B = S-P.
\]

We now compare Eq.~\eqref{eq:standard_replicator} with the strategy dynamics of the deterministic system,
\begin{equation}
\frac{dx}{dt} = \nu\,x(1-x)\bigl(1-2\rho'\bigr),
\end{equation}
where \(\rho'\) denotes the rescaled resource level.
In general, since \(\rho'\) evolves dynamically and depends on \(x\), the strategy dynamics cannot be written as a standard replicator equation with constant payoff entries.
However, when the resource relaxes to a non-null stationary state \(\rho=\rho^*\), defined by
\begin{equation}
\rho^* = 1-\frac{g(x)}{b},
\end{equation}
where $g(x)$ is a function solely of $x$, the strategy equation can be evaluated at this stationary resource level.
Substituting \(\rho^*\) into the strategy dynamics yields
\begin{equation}
\frac{dx}{dt}
= \nu\,x(1-x)\!\left(1-2\frac{1-\frac{g(x)}{b}}{1-\frac{e_{cc}}{b}}\right).
\end{equation}
After straightforward algebraic manipulations, this equation can be rewritten as
\begin{equation}
\frac{dx}{dt}
= \tilde{\nu}\,x(1-x)\bigl(Qx^2 + Ax + B\bigr),
\label{eq:nonstandard_replicator}
\end{equation}
where
\[
Q = \frac{2}{b}(e_{cc}-e_{cd}-e_{dc}+e_{dd}), \qquad
A = \frac{2}{b}(e_{cd}+e_{dc}-2e_{dd}),
\]
\[
B = \frac{2e_{dd}}{b}-\frac{e_{cc}}{b}-1, \qquad
\tilde{\nu}=\frac{\nu}{1-\frac{e_{cc}}{b}}.
\]

Equation~\eqref{eq:nonstandard_replicator} has the same functional form as the standard replicator equation~\eqref{eq:standard_replicator} only in the special case \(Q=0\). The condition $Q=0$ corresponds to a symmetric scenario in terms of per-capita extraction rates. In this case, the sum of the harvesting rates in a fully cooperative population $(e_{cc})$ and a fully defective population $(e_{dd})$ is exactly equal to the sum of the harvesting rates in a mixed population $(e_{cd}+e_{dc})$. Hence $e_{cc}+e_{dd}=e_{cd}+e_{dc}$, in other words, the total exploitation of the resource is balanced: the combined effect of the extreme populations (all cooperators or all defectors) produces the same total resource extraction as a population composed of a mix of strategies.

The parameter choice adopted in the main text satisfies this condition, enabling a consistent mapping to an effective payoff matrix.
Under this assumption, the coefficients of the replicator equation can be identified as
\[
R-S-T+P =\frac{2}{b}(e_{cd}+e_{dc}-2e_{dd}),
\qquad
S-P =  \frac{2e_{dd}}{b}-\frac{e_{cc}}{b}-1.
\]
Since these relations provide two constraints for four payoff entries, two parameters remain free.
Choosing \(S\) and \(T\) as independent parameters, the effective payoff matrix can be written as
\begin{equation}
\Pi_{\mathrm{eff}} =
\begin{pmatrix}
\frac{e_{cc}}{b}-1+T & S \\[2mm]
T & S+1+\frac{e_{cc}}{b} -\frac{2e_{dd}}{b}
\end{pmatrix}.
\end{equation}
\subsection{Classification of game regimes}
We characterize the game-theoretic structure induced by the effective payoff matrix $\Pi_{\mathrm{eff}}$ through the invasion quantities $T-R$ and $P-S$, whose signs fully determine the phase diagram of the replicator dynamics. From the effective payoff matrix $\Pi_{\mathrm{eff}}$, these quantities are given by:

\begin{align}
T - R &= 1-\frac{e_{cc}}{b}, \\
P - S &= 1 - 2\frac{e_{dd}}{b} + \frac{e_{cc}}{b}.
\end{align}

The first difference, $T-R$, quantifies the temptation to defect when all other players are cooperating. 
Because the model is defined on the domain $e_{cc} < b$, this difference is strictly positive ($T - R > 0$).
This allows us to rule out game classifications characterized by a negative $T-R$ value, such as coordination and harmony games (Stag-Hunt and Harmony games). 

As a result, the system is restricted to two game classes: the Chicken game and the Prisoner's Dilemma. The parameter space and are separated by the nullcline $P-S=0$, which yields the boundary $e_{dd} = \frac{e_{cc}+b}{2}$.  
Specifically the game is classified as a Chicken Game when $e_{dd} > \frac{e_{cc}+b}{2}$, which is always satisfied for $e_{dd}>b$, and the Prisoner Dilemma otherwise.
Hence the Chicken game regime is always accessible and as pointed out by Ito~\cite{ito_complete_2024}, when modeling shared common resources, the Chicken game often emerges as the more natural framework. In this context, an abundance of resources incentivizes defection; conversely, when resources are scarce and all other players defect, mutual defection results in the worst possible outcome.

\end{document}